\documentstyle[11pt,aspconf]{article}
\begin{document}

\def\s{{\it 2nd-P~}}
\def\ss{{\it 2nd-Ps~}}
\def\etal{{\it et al.~}}
\def\y{{\it Yale-group}}
\def\r{$R_{GC}$}
\def\kms{{km s$^{-1}$}}
\def\fe{{$[Fe/H]$ }}
\def\lr{$Log R_{GC}$ }
\def\ros{$\rho_0$ }
\def\ro{$log \rho_0$ }
\def\rc{$r_c$ }
\def\age{$Age_{TO}$ }
\def\agec{$Age_{HB}$ }
\def\mv{$|M_V|$ }
\def\si{$\sigma_0$ }
\def\bv{$(B-V)_{peak}$ }
\def\bu{$B2-R/B+R+V$ }
\def\b{$B2/B+R+V$ }
\def\zi{$B-R/B+R+V$ }
\def\vs{{\it vs.} }
\def\ie{{\it i.e.} }

\title{Horizontal Branch Morphology and the 2nd Parameter Problem}

\author{F. Fusi Pecci, M. Bellazzini, F.R. Ferraro}
\affil{Osservatorio Astronomico, Bologna, Italy}
\author{R. Buonanno, C.E. Corsi}
\affil{Osservatorio Astronomico, Roma, Italy}

\begin{abstract}
We review the most outstanding issues related to the study of the morphology
of the Horizontal Branch (HB) in the Color--Magnitude Diagrams
of Galactic Globular Clusters and its use as {\it age indicator}.
It is definitely demonstrated (see also Bolte, this meeting) that age
cannot be {\it the only} \s driving the HB morphology.
Other candidate \ss are briefly examined, with special attention
to the possible influence of cluster stellar density.

A procedure is presented through which the measured HB morphology can 
be depurated from the effects of parameters other than age so that a 
rough determination of the age of globulars could still be obtained 
from their HBs. This method could be very useful for dating globular 
clusters in nearby galaxies.
\end{abstract}

\section{Introduction}
The Horizontal Branch (HB) stars have almost the same absolute luminosity at 
the same evolutionary stage, and this makes them (RR Lyrae variables) 
fundamental distance indicators. Moreover, 
as a result of large differences in the radius, HB stars present a
wide color (temperature) distribution, usually called HB morphology,
which is a unique astrophysical tool potentially, but
whose complete understanding is still matter of continuous debate.

In general globular cluster HBs get redder as metallicity increases.
Besides metallicity, [Fe/H] --the {\it first} parameter-- the
classification of HB morphologies requires however the introduction of a still
unidentified {\it second} parameter ({\it 2nd-P}) as perceived during
the early 1960's by Sandage and Wallerstein (1960), Faulkner (1966),
van den Bergh (1967), and Sandage and Wildey (1967). 

\section {Why does the HB morphology vary so greatly between clusters of similar
composition?} 
This is the essence of the so-called ``\s problem'' which
nowadays embraces the whole subject related to studying HB morphologies.
Hundreds of papers have somehow dealt with this hot subject 
during the last thirty years. Useful discussions and references are
reported for instance by Renzini (1977), Kraft (1979), Freeman and Norris 
(1981), Buonanno \etal (1985), Zinn (1986), Rood and Crocker (1989), 
Bolte (1989), VandenBerg \etal (1990), Sarajedini and Demarque (1990),
Fusi Pecci \etal (1993), Buonanno (1993), Lee (1993), 
van den Bergh (1993), Catelan and de Freitas Pacheco (1994), 
Lee \etal (1994), Peterson \etal (1995).

Many candidate \ss have been proposed including stellar parameters and 
global cluster properties, and in particular:
age --$t$--, primordial He abundance --Y$_p$-- (but dredged-up Helium
might also be important), CNO abundance --[CNO/Fe]--, 
rotation --$\omega$ ~~with $\omega \uparrow \Rightarrow $ mass 
loss$\uparrow$--, cluster central concentration and density --\ros
~~with \ros $\uparrow \Rightarrow $ mass loss $\uparrow$ (see Fig. 1).
To summarize briefly the major effects,
we report in Fig. 1 a few indicative plots which can be used as a sort 
of ``tourist map'' in the \s game.

Unfortunately, models addressing the \s require assumptions about
stellar mass loss which is one of the most crucial events during the 
post-main-sequence evolution leading to the HB stage (Rood 1973).
It is because of our too deep ignorance about mass
loss and because of these assumptions that we have been less {\it
optimistic} than others (Lee \etal 1990, 1994) in identifying {\it the}
\s as age. Indeed, the physical parameter which varies from star
to star leading to the differential mass loss commonly thought to cause
the HB color distribution {\it has never been identified}. We firmly believe
that the complete solution of the so-called \s problem can hardly be achieved 
without identifying the parameter which leads to differential
mass loss within a cluster.

\section{Why is the identification of the \s so important?}
One can quote two main reasons, at least. 

First, given the exceptional sensitivity of HB star 
location in color to almost any model parameter (Rood 1973), a complete
understanding of the HB morphology yields a detailed test of the
results of evolution theory of Pop II stars (Renzini and Fusi Pecci 1988).

Second, since evidence has 
mounted (Bolte 1989, Green ans Norris 1990, Sarajedini and
Demarque 1990, VandenBerg, Bolte and Stetson 1990, Dickens \etal 1991,
but see Catelan and de Freitas Pacheco 1994)
that the \s clusters NGC 362 and NGC 288 have strongly different
HB morphologies because of an age difference, justification has grown
for the use of HB morphology as a powerful {\it relative} age indicator.

Although the first item is surely not less important, the attention
has been mostly centered on the second aspect, mainly because
any route to get a ``calibrated clock'' is crucial to astronomy.

For instance,
the overall description of the globular cluster system and
early stages of Galaxy evolution proposed by the {\it Yale-group}
(Zinn 1986, 1993, Lee \etal 1990, 1994, Lee 1993, Armandroff 1993) 
starting from the seminal work by Searle and Zinn (1978) adopts 
the HB morphology as ``bona fide clock'',
and the present Galactocentric location and radial velocity, as indicators 
of the kinematical properties.

Once age has been justified as {\it the} \s, 
one can use this result also to show for instance
that the mean age of the field HB stars decreases outward in the halo
(Preston \etal 1991, Suntzeff \etal 1991) 
or to try to identify which clusters (or group
of clusters) belong to a given original ``fragment'' (Rodgers and 
Paltoglou 1984, Zinn 1993, van den Bergh 1993, 1994, Majewski 1994).

\begin{figure}
\epsscale{0.8}
\plotone{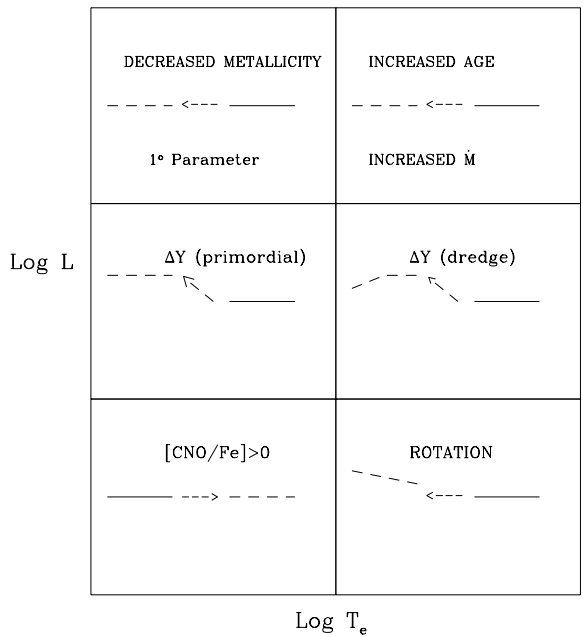}
\caption{The \s game: a "tourist map". Blue to the left.}
\end{figure}

\section{HB morphology: a ``bona fide'' clock?}
When one looks in details, the ``HB morphology world'' is indeed very complex
as one finds gaps, bimodalities, blue tails, radial population gradients, etc..
Therefore, we and others (see Buonanno \etal 1985, Fusi Pecci 
\etal 1993, 1995 for references) have posed the question: {\it  is
age the only} \s? If not, the definition of ``young'' based on
the HB morphology may be misleading or, at least, it could be 
different from that based on the {\it actual
measure} of the Main Sequence Turnoff (TO). This difference may well be
irrelevant eventually, but we must investigate further the problem.
Note that in the past few years a small set of young (from the TO) globular 
clusters has been detected (Pal 12, Ruprecht 106, Arp 2, Terzan 7, and 
IC 4499, see for references Fusi Pecci \etal 1995). They are peculiar
in many respects and could now belong to the Milky Way as a result of capture,
but can offer some useful hints in the study of the \s problem.

\section{Global versus non-global \ss}
Following Freeman and Norris
(1981), it is nowadays quite common to discuss about the so-called
{\it global} and {\it non-global} \ss. The difference between the
two groups is clearly explained in the following statement from Lee (1993):
``...some HBs apparently have complex structures, which seem to require a 
\s internal to those peculiar clusters. I will refer to those features as
the {\it non-global} \s phenomenon. Although they are clearly important
to our understanding of the details of the HB and probably also the
structure of globular clusters, they appear to have minor effect
on the {\it global} \s phenomenon, which I define here as the mean
systematic variation of HB morphology with Galactocentric distance
--$R_{GC}$''.

The \y ~(Lee \etal 1994) has carefully considered the possible effects 
of the \ss proposed so far, {\it i.e.} $t$, Y$_p$, [CNO/Fe], $\omega$,
\ros. 
They concluded that (1) ``the variations in Y$_p$, [CNO/Fe], and core mass 
(in turn, $\omega$) required separately to explain the {\it global} \s
phenomenon are inconsistent  with either the observed properties
of the RR Lyrae variables or the observed MS-TOs in the clusters'';
(2) ``there is no clear evidence that the trend with \r ~is related to
\ros or absolute magnitudes of the clusters''; (3) ``the variations
in cluster age required to explain the trend with \r ~are not in conflict
with any observations''. Hence {\it age is the global} \s. 

\section{Are the effects of non-global \ss always present and important?}
A typical example of the fact that there are still outstanding questions
remaining is for instance the case of NGC 2808. The HB of NGC 2808
is bimodal (Ferraro \etal 1990), and in fact resembles a superposition 
(Rood \etal 1993) of the HBs of NGC 362 and NGC 288, the two clusters
which suffer the \s syndrome in the opposite direction and which have
been adopted as ``milestone'' in the claim: {\it age is the} \s
(see Lee 1993 Fig. 6  and Bolte 1993, Fig. 1 compared to Rood \etal 1993,
Fig. 1). 
Since the \s affecting the two groups of HB stars in NGC 2808 is not likely
to be age (a 4-6 Gyr difference would be necessary!), one is forced to
conclude that (at least in this specific case) we must consider a {\it
third} parameter. 

Moreover, the detection of peculiar features (in clusters 
at any \r) is becoming a rule rather than a specific exception, and
the latest HST data on the distant cluster Pal 3 (see Bolte, this meeting)
shows that it is ``old'' and with a red HB. Hence, also
the subdivision itself of {\it global} and {\it non-global} 
quantities becomes dialectic. 

Consequently, one has to split the \s problem into two distinct parts:
the first, to study which are the candidate \ss that mostly affect the 
location of the bulk of the HB star distribution; the second, to understand
which are those able to alter the HB distribution, leading to bimodalities,
gaps, blue tails, etc., and to determine quantitatively how important are 
the relative contributions to the actual observed HB morphology in each 
observed cluster.

For these reasons we are reluctant to believe that age, or any other
single parameter, could be {\it the} \s. In our view, age is probably 
{\it one of the many } \ss, possibly the most important. As suggested 
by Buonanno \etal (1985), the definition of the {\it global} parameter has to 
be extended to mean that the parameter driving the observed HB morphology
is actually the result of the ``global combination'' of many individual
quantities and phenomena related to the formation and the chemical and
dynamical evolution of each individual star in a cluster and of each
cluster as a whole. 

Within this framework, it is important to take into account the growing body
of observational results suggesting the existence of a possible link between 
the dynamical history of globular clusters and the evolution of their stellar
members (including blue stragglers, binaries, etc.; they {\it are} indeed 
cluster members! see for references Fusi Pecci \etal 1992, Bailyn 1995). 
In particular, Fusi Pecci \etal (1993, hereafter FPAL) have
shown that the color extension of the HB distribution is correlated 
with the cluster central density --\ros, in the sense that more 
centrally concentrated clusters tend to have bluer HB-morphologies 
and often present extended blue tails. 
Since total luminosity and central density are correlated (Djorgovski and 
Meylan 1994), van den Bergh and Morris (1994) noted that the 
correlation found by FPAL could be a spurious effect reflecting the 
cluster-to-cluster variation in total luminosity. 

In the following sections we report schematically the results of a study 
in progress (Buonanno \etal 1996, hereafter BAL) which support FPAL and 
show that the cluster density could even be a {\it global} \s in the sense 
defined by Lee (1993). 
Finally, we show that the simple use of HB-morphology as age indicator 
without taking into account other possible effects could not be appropriate,
in particular for the clusters of intermediate metallicity ($-1.8\le 
[Fe/H]\le -1.4$) where the HB morphology displays its maximum sensitivity
to any variation.

\begin{figure}
\epsscale{0.65}
\plotone{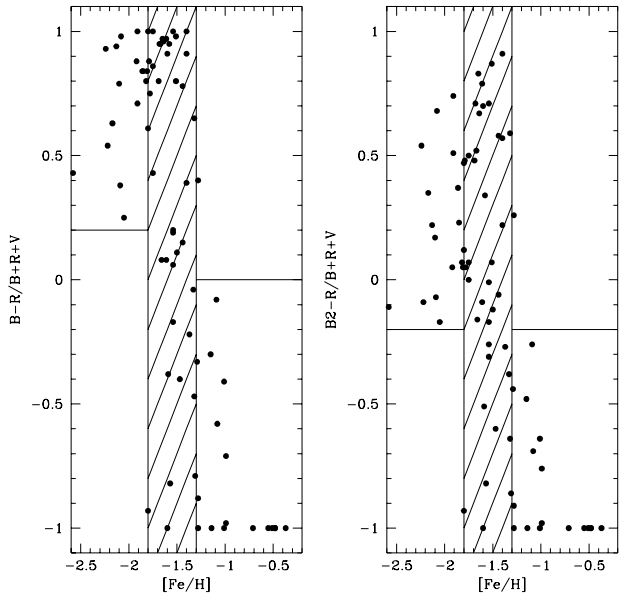}
\caption{Observables describing the HB morphology \vs metallicity.
Note the large variation of the HB morphology in the intermediate-metallicity
regime (shaded area)}
\end{figure}

\section{The new HB-morphology observables and the data-base}
One of the main difficulties in studying the HB consists in the difficulty to 
yield a satisfactory quantitative description of its detailed morphology. 
The \y ~(Zinn 1986, Lee 1990) defined the observable \zi, where V, B, R 
are the number of stars located in the CMD within the instability strip, 
to its blue and red, respectively. This index works like a ``wide-band 
HB-morphology photometer''. It is easy to compute, but has a ``low
resolution'' and ``saturates'' for very blue (\zi=1) and very red 
(\zi=-1) HBs. FPAL have shown that it is mainly sensitive to the position 
of the peak of the HB distribution.

To describe the blue side of the HB distribution, Preston \etal (1991) 
defined the index $B_W$, that is the number of HB stars in the color 
interval $-0.02<(B-V)_0<0.18$. Later, Buonanno (1993) adopted 
\bu ~having defined $B2$ as the complement of $B_W$ to $B$,
{\it i.e.} $B2=B-B_W$. While the clusters with blue HB tails
saturates in the interval $0.9<B-R/B+R+V\le 1.0$, \bu spans
a much larger interval. For instance, Arp 2, NGC 5897, NGC 6341 and NGC 6218 
(with highly different HBs)  have $B-R/B+R+V\sim 0.88$, whereas \bu 
ranks them with $-0.22, 0.09, 0.45$ and $0.77$, respectively.
To measure quantitatively the blue tails we also defined \b, {\it i.e.} the 
{\it normalized} number of stars located in the HB blue tail (to overcome the
the point raised to FPAL by van den Bergh and Morris 1994).

Our complete database collects data for 69 halo globular clusters for which
we have obtained the above HB-morphology observables from the available
photometries. \fe, \ro and \r ~are taken from Djorgovski (1993). The de-reddened 
colors of the peak of the HB-distributions --$(B-V)_{peak}$-- are drawn 
from FPAL or newly computed. The ``relative'' ages based on TO-properties
for 27 clusters are taken from BAL, as well as the
whole estimate and treatment of the errors. Note that the use of other
data-sources and the ages recently presented by Chaboyer \etal (1995)
do not modify the essence of these results (see BAL for a discussion).

As shown by FPAL (Fig. 1), the sensitivity of the
location in temperature (color) of the HB stars is dependent
on metallicity in a strongly non-linear way. Therefore, we selected
from the plots shown in Fig. 2 the region of maximum sensitivity,
$-1.8\le [Fe/H]\le -1.4$, and present here just the results for this
intermediate-metallicity window.

\begin{figure}
\epsscale{0.60}
\plotone{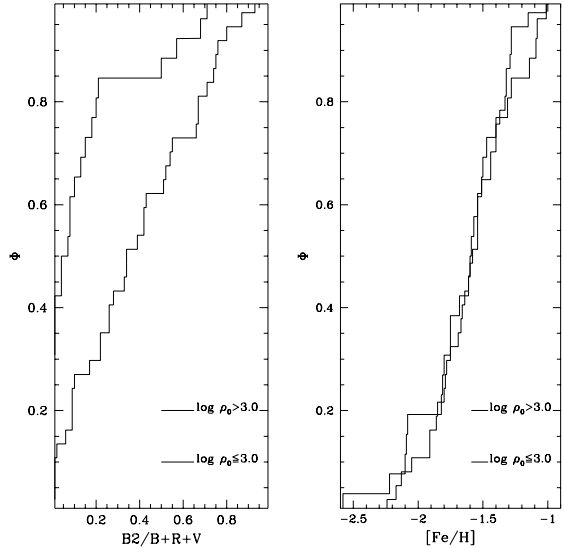}
\caption{Cumulative distributions in $B2-R/B+R+V$ (left panel) and in $[Fe/H]$
(right panel) for clusters of different central densities}
\end{figure}

\section{Correlations between \b and \ro and \r}
In order to check the possible connection between \ro and HB morphology
claimed by FPAL, we divided the sample in two groups with:
(i) $log \rho_0>3$ (39 objects) and (ii) $log \rho_0\le 3$ (26 objects).
As shown in Fig. 3, the two cumulative distributions with respect to
\b are strongly different, whereas they are indistinguishable with respect
to [Fe/H]. A KS-test yield a probability of only $0.02 \%$ that the two 
groups are extracted from the same population in \b. Hence, if \ros
$\uparrow ~~\Rightarrow$ HB morphology gets bluer.

Due to shortage of space, a complete quantitative analysis and demonstration
on how cluster stellar density may play a r\^ole in influencing the
HB morhology can be found in FPAL and in the the forthcoming paper (BAL). 
However, once 
demonstrated the existence of this significant influence, at least for 
intermediate-metallicity clusters (where HB sensitivity
is maximum), and considering that nearly half of the halo clusters are of
intermediate metallicity, it would be quite evident that using HB-morphology as 
age indicator without taking into account the possible effects due to
the differences in the cluster central density may produce 
very noisy, if not misleading, age scales even if age were the
{\it dominant} \s.

In particular, looking at the mean systematic variation of HB morphology
with \r ~(the {\it global} \s effect), it can be argued 
that this phenomenon could be partially due to the well known anti-correlation
existing between \ro and \r ~coupled with the tendency of 
loose clusters to display redder HB morphologies (at fixed metallicity) 
as claimed above.

\begin{figure}
\epsscale{0.98}
\plotone{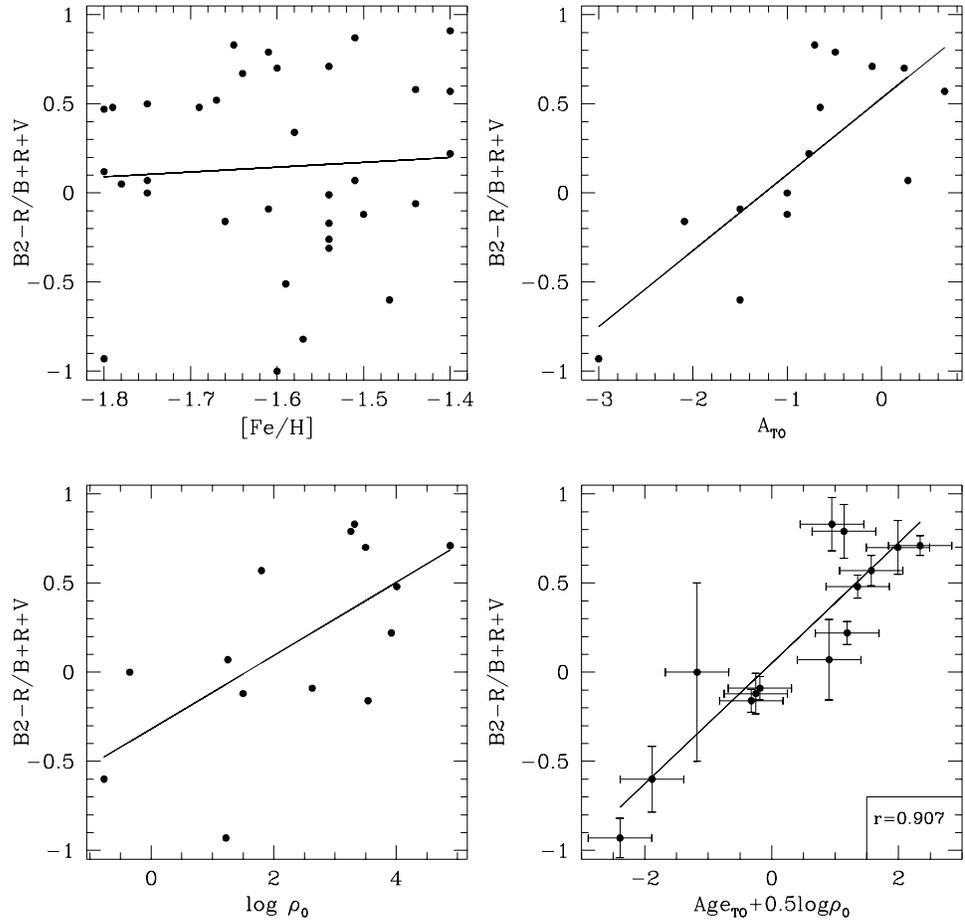}
\caption{The dependence of $B2-R/B+R+V$ on $[Fe/H]$, $A_{TO}$ and $log \rho_0$
for the clusters with $-1.4\le [Fe/H]\le -1.8$. In the bottom-left panel the 
bivariate correlation is shown.}
\end{figure}

\section{Playing with HB morphologies and ages}
Having at disposal a set of independent {\it relative} cluster ages,
one can play with the data to
verify whether, under the assumption that age is {\it not} the only \s,
a proper combination of the effects due to age and, for instance,
to differences in \ros can offer a more stringent correlation with
the observable adopted to describe the HB morphology.

Due to the poorness of the sample and the still high uncertainties on 
the relative ages (here expressed as differences with respect to an
arbitrary zero-point adopted as reference by BAL), the results 
presented in Fig. 4 are essentially indicative. However, we find
a curious indication, at least.

In synthesis, Fig. 4a shows the distribution of the clusters having
$-1.8\le [Fe/H]\le -1.4$ in the plane \bu \vs [Fe/H]. As can be seen,
within this metallicity interval, where the sensitivity of the 
HB morphology to the variation of any parameter is highest, there is
no correlation with [Fe/H] --the {\it first} parameter. Panel {\it b}
shows that, for the few clusters for which TO-ages are available,
there is a quite evident correlation with \bu. In addition, 
panel {\it c)} shows that a (slightly weaker) correlation can be found 
for the same clusters also between \bu ~and \ro. 
Finally, panel {\it d} leads to the most
interesting evidence that, if the effects of the two possible 
dependences of the HB morphology are combined via a bivariate analysis 
(\ie by seeking for the best linear combination of \age and \ro which 
optimizes the ranking of the \bu values), one obtains a significant improvement
of the correlation with respect to the monovariate cases (panels {\it
b,c}). The linear correlation coefficient yields a probabiblity less
than $0.05\%$ that the two vectors in panel {\it d} are actually
uncorrelated.

Based on this exercise, which of course requires a much deeper
analysis and confirmation, one might conclude that the manifold
of HB morphologies at fixed metallicity is at least {\it two-dimensional}.
In the overall scenario of the classification of the HB morphology
of Galactic GC's, this would imply that while metallicity is {\it the first} 
parameter, {\it the} \s is probably a complex combination of the
effects due to various \ss, or, if one wants to keep the major r\^ole
associated to age, that a complete understanding of the HB morphology
requires the considerations of a {\it third} parameter, at least.


\end{document}